\documentclass[prl,twocolumn,superscriptaddress]{revtex4}  
\usepackage{graphicx}  
\usepackage{dcolumn}   
\usepackage{bm}        
\usepackage{amssymb}   
\usepackage{amsmath}

\def\beq{\begin{equation}}
\def\eeq{\end{equation}}
\newcommand{\bea}{\begin{eqnarray}}
\newcommand{\eea}{\end{eqnarray}}

\def\bsp#1\esp{\begin{split}#1\end{split}}

\newcommand{\eps}{\epsilon}

\newcommand{\cI}{\begin{cal}I\end{cal}}

\newcommand{\cS}{{\cal S}}

\newcommand{\rd}{\mathrm{d}}

\newcommand{\mbint}{\int_{-i\infty}^{+i\infty}}

\def\bit#1\eit{\begin{itemize}#1\end{itemize}}
\def\ben#1\een{\begin{enumerate}#1\end{enumerate}}

\newenvironment{sloppyequation}[0]{\sloppy\begin{flushleft}\hspace*{0.75cm}\(}{\)\end{flushleft}\fussy}
\newenvironment{sloppytext}[0]{\sloppy\begin{flushleft}}{\end{flushleft}\fussy}

\newcommand{\beqsloppy}{\begin{sloppyequation}}
\newcommand{\eeqsloppy}{\end{sloppyequation}}
\newcommand{\btxtsloppy}{\begin{sloppytext}}
\newcommand{\etxtsloppy}{\end{sloppytext}}

\begin{document}

\mbox{DCPT/11/34, IPPP/11/17}

\author{Vittorio Del Duca}
\affiliation{INFN, Laboratori Nazionali Frascati, 00044 Frascati (Roma), Italy,\\ Email: {\tt delduca@lnf.infn.it}}

\author{Claude Duhr}
\affiliation{Institute for Particle Physics Phenomenology,
University of Durham, Durham, DH1 3LE, U.K.,\\ Email: {\tt claude.duhr@durham.ac.uk}}

\author{Vladimir A. Smirnov}
\affiliation{Nuclear Physics Institute of Moscow State University,
Moscow 119992, Russia,\\ Email: {\tt smirnov@theory.sinp.msu.ru}}

\title{The massless hexagon integral in $D=6$ dimensions}

\begin{abstract}
We evaluate the massless one-loop hexagon integral in six dimensions. The result is given in terms of standard polylogarithms of uniform transcendental weight three, its functional form resembling the one of the remainder function of the two-loop hexagon Wilson loop in four dimensions.

\end{abstract}

\maketitle


In this short note we are concerned with the computation of the scalar one-loop integral in $D=6$ dimensions,
\beq\label{eq:massless_hexagon}
I_6^{D=6} = \int{\rd^6k\over i\pi^3}\prod_{i=0}^5{1\over D_i}\,,
\eeq
with 
\beq
D_0 = k^2 {\rm~~and~~} D_i = (k+p_i)^2, {\rm~~for~~} i=1,\ldots,5\,.
\eeq

The external momenta, labeled by $p_i$, $i=1,\ldots,6$, are lightlike, $p_i^2=0$, and all ingoing, such that momentum conservation reads
\beq
\sum_{i=1}^6 p_i=0\,.
\eeq
We consider the integral in Euclidean kinematics where all Mandelstam invariants are taken to be negative, $(p_1+\ldots+p_j)^2<0$, and the integral is real.
The massless hexagon integral is finite in $D=6$ dimension, so that no regularization is required and we can perform the computation in strictly six dimensions. 

We introduce dual coordinates~\cite{Drummond:2006rz,Alday:2007hr},
\beq
p_i = x_i - x_{i+1}\,,
\eeq
with $x_7=x_1$, due to momentum conservation.
Since the integration measure in Eq.~\eqref{eq:massless_hexagon} is translation invariant, we can define $k=x_0-x_1$ and the integral can be rewritten completely in terms of dual coordinates,
\beq\label{eq:massless_hexagon_X}
I_6^{D=6} = \int{\rd^6x_0\over i\pi^3}{1\over x_{01}^2\,x_{02}^2\,x_{03}^2\,x_{04}^2\,x_{05}^2\,x_{06}^2}\,,
\eeq
with $x_{ij}^2 = (x_i-x_j)^2 = (p_i+\ldots+p_{j-1})^2$. In Ref.~\cite{Drummond:2006rz} the notion of dual conformal invariance was introduced, \emph{i.e.}, the action of the conformal group on the dual coordinates $x_i$. The integral~\eqref{eq:massless_hexagon_X} transforms covariantly under dual conformal transformations. In fact, invariance under rotations and translations is manifest, whereas under dilatations and inversions the integral transforms covariantly with weight one at each external point $x_i$, namely under dilatations, $x_i\to \lambda\,x_i$, the integral scales as $I_6^{D=6}\to \lambda^{-6}\,I_6^{D=6}$, whereas under inversions $x_i\to x_i/(x_i^2)^2$ the measure and the propagators transform as $\rd^6x_0\to \rd^6x_0/(x_0^2)^6$ and $x_{ij}^2\to x_{ij}^2/(x_i^2x_j^2)$, such that $I_6^{D=6}\to I_6^{D=6}\,\prod_{i=1}^6 x_i^2$. Note that for dual conformal invariance to hold it is crucial that we work in strictly six dimensions. Finally, the previous considerations are not restricted to $I_6^{D=6}$, but exactly the same reasoning shows that every finite one-loop $n$-gon in $D=n$ dimensions is dual conformally covariant.

A direct consequence of the dual conformal covariance of $I_6^{D=6}$ is that the integral can only depend on dual conformal cross ratios, up to an overall prefactor which carries the conformal weights. For the massless six-point kinematics, there are three independent cross ratios~\cite{Drummond:2007bm}, given in terms of dual coordinates by,
\beq
u_1 = {x_{15}^2\,x_{24}^2\over x_{14}^2\,x_{25}^2}\,,\,\,\,
u_2 = {x_{26}^2\,x_{35}^2\over x_{25}^2\,x_{36}^2}\,,\,\,\,
u_3 = {x_{31}^2\,x_{46}^2\over x_{36}^2\,x_{41}^2}\,.
\eeq
More precisely, the integral can be written in the form
\beq
I_6^{D=6} = {1\over x_{14}^2\,x_{25}^2\,x_{36}^2}\,\cI_6(u_1,u_2,u_3)\,.
\eeq
where the function $\cI_6(u_1,u_2,u_3)$ is manifestly dual conformal invariant. Furthermore, the integral $I_6^{D=6}$ as a function of the external momenta $p_i$ has a dihedral symmetry $D_6$ generated by cyclic rotations $p_i\to p_{i+1}$ and  the reflection $p_i\to p_{6-i+1}$. It is easy to check that the dihedral symmetry of $I_6^{D=6}$ implies that the function $\cI_6(u_1,u_2,u_3)$ must be totally symmetric in the three cross ratios.

We start by deriving a Mellin-Barnes (MB) representation for $I_6^{D=6}$ using the {\tt AMBRE} package~\cite{Gluza:2007rt}. Although the integral is finite, the resulting MB representation has a spurious singularity that must cancel in the end. We therefore derive the MB representation in $D=6-2\eps$ dimensions and resolve the singularities in $\epsilon$ using the strategy introduced in Refs.~\cite{Smirnov:1999gc,Tausk:1999vh,Smirnov:2004ym,Smirnov:2006ry} by applying the codes
 {\tt MB}~\cite{Czakon:2005rk} and {\tt MBresolve}~\cite{Smirnov:2009up}
and obtain a set of MB integrals which can be safely expanded in $\eps$ under the integration sign. After applying these codes, all the integration contours are straight vertical lines. At the end of this procedure, all the poles in $\epsilon$ cancel and we are left with a manifestly finite and conformally invariant threefold MB integral to compute,
\beq\bsp\label{eq:MB_integral}
\cI_6 &\,= \mbint\left(\prod_{i=1}^3{\rd z_i\over 2\pi i}\,\Gamma(-z_i)^2\,u_i^{z_i}\right)\\
&\,\times\Gamma(1+z_1+z_2)\,\,\Gamma(1+z_2+z_3)\,\Gamma(1+z_3+z_1)\,,
\esp\eeq
where the contours are straight vertical lines whose real parts are given by
\beq
\textrm{Re}(z_1) = -{1\over3}\,,\,\,
\textrm{Re}(z_2) = -{1\over4}\,,\,\,
\textrm{Re}(z_3) = -{1\over5}\,.
\eeq
Albeit simpler, the integral~\eqref{eq:MB_integral} is similar to the threefold MB integral contributing to the two-loop hexagon Wilson loop in $\begin{cal}N\end{cal}=4$ Super Yang-Mills~\cite{DelDuca:2009au, DelDuca:2010zg}, hence it can be computed in the same fashion. Following the strategy of Ref.~\cite{DelDuca:2010zg}, we can turn each MB integration into an Euler-type integral via the formula,
\beq\bsp\label{eq:MBtoEuler}
&\mbint{\rd z\over 2\pi i}\,\Gamma(-z)\,\Gamma(c-z)\,\Gamma(b+z)\,\Gamma(c+z)\,X^{z}\\
&=\Gamma(a)\,\Gamma(b+c)\,\int_0^1\rd v\,v^{b-1}\,(1-v)^{a+c-1}\,(1-\overline{X}v)^{-a}\,,
\esp\eeq
with $\overline{X} = 1-X$ and where the contours are such as to separate the poles in $\Gamma(\ldots-z_i)$ from those in $\Gamma(\ldots+z_i)$. This leaves us with the following three-fold parametric integral to compute,
\beq\bsp
&\cI_6=\int_0^1\left(\prod_{i=1}^3\rd v_i\right)
\frac{1}{1-v_2 \left(1-u_1 v_1\right)}\\
&\times\frac{1}{1-v_1 \left(1-u_2-v_3 \left(1-u_2-u_3 v_2\right)\right)-(1-u_3 v_2) v_3}\,.
\esp\eeq
The integral is easily performed in terms of multiple polylogarithms~\cite{Goncharov:1998}. The resulting expression is rather lengthy and involves a combination of multiple polylogarithms of uniform weight three, whose arguments are complicated algebraic functions involving the square root of the quantity,
\beq\label{eq:Delta}
\Delta = (u_1+u_2+u_3-1)^2-4u_1u_2u_3\,.
\eeq
However, the similarity between the MB integral~\eqref{eq:MB_integral} and the corresponding integral of Ref.~\cite{DelDuca:2010zg} suggests that it should be possible to rewrite the answer in a simpler form, in the same way as the analytic result of Ref.~\cite{DelDuca:2010zg} was rewritten in simplified form in Ref.~\cite{Goncharov:2010jf}. The cornerstone of the simplification of the two-loop six-point remainder function was the so-called symbol map, a linear map $\begin{cal}S\end{cal}$ that associates a certain tensor to an iterated integral, and thus to a multiple polylogarithm. In the following we give a very brief summary of the symbol technique, referring to Ref.~\cite{Goncharov:2010jf} for further details. As an example, the tensor associated to the classical polylogarithm $\textrm{Li}_n(x)$ is,
\beq
\begin{cal}S\end{cal}(\textrm{Li}_n(x)) = -(1-x)\otimes\underbrace{x\otimes\ldots\otimes x}_{(n-1) \textrm{ times}}\,.
\eeq
Furthermore, the tensor maps products that appear inside the tensor product to a sum of tensors,
\beq
\ldots\otimes(x\cdot y)\otimes\ldots = 
\ldots\otimes x\otimes\ldots+
\ldots\otimes y\otimes\ldots\,.
\eeq
It is conjectured that all the functional identities among (multiple) polylogarithms are mapped under the symbol map $\cS$ to algebraic relations among the tensors. Hence, if the symbol map is applied to our complicated expression for $\cI_6(u_1,u_2,u_3)$, it should capture and resolve all the functional identities among the polylogarithms, and therefore allow us to rewrite the result in a simpler form.
In order to apply this technology, it is however important that all the arguments that enter the tensor be multiplicatively independent. As in our case the arguments of the polylogarithms involve square roots of $\Delta$, this requirement is not fulfilled. In Ref.~\cite{Goncharov:2010jf} a reparametrization of the cross ratios $u_i$ in terms of six points $z_i$ in $\mathbb{CP}^1$ was proposed,
\beq\label{eq:zij}
u_1 = {z_{23}z_{56}\over z_{25}z_{36}}\,,\,\,
u_2 = {z_{34}z_{61}\over z_{36}z_{41}}\,,\,\,
u_3 = {z_{45}z_{12}\over z_{41}z_{52}}\,,
\eeq
with $z_{ij} = z_i - z_j$. It is easy to check that with this parametrization the right-hand side of Eq.~\eqref{eq:Delta} becomes a perfect square. Hence, after this reparametrization all the arguments of the polylogarithms are rational functions in the $z_{ij}$ variables, making this parametrization well suited to apply the symbol map $\cS$.

Using the parametrization~\eqref{eq:zij} and the symbol map, it is easy to construct a simpler candidate expression with the same symbol as our original expression. However, the kernel of the map $\cS$ is non trivial, and it allows us to fix the candidate expression only up to terms proportional to zeta values, which in turn must be fixed by looking at particular values of the cross ratios.
At the end of this procedure, we arrive at the following expression for the scalar massless hexagon integral,
\beq\bsp\label{eq:result}
\cI_6&(u_1,u_2,u_3) = {1\over\sqrt{\Delta}}\,\Bigg[
-2\sum_{i=1}^3L_3(x_i^+,x_i^-) \\
&\,+ {1\over3}\left(\sum_{i=1}^3\ell_{1}(x_i^+) - \ell_{1}(x_i^-)\right)^3\\
&\,+{\pi^2\over 3}\,\chi\,\sum_{i=1}^3(\ell_{1}(x_i^+) - \ell_{1}(x_i^-))\Bigg]\,,
\esp\eeq
where
\beq
x_i^\pm = u_i\,x^\pm\,,\quad x^\pm = {u_1+u_2+u_3-1\pm\sqrt{\Delta}\over 2u_1u_2u_3}\,,
\eeq
and $\Delta$ is defined in Eq.~\eqref{eq:Delta}, and with
\beq\bsp
L_3&(x^+,x^-)\\
&=\sum_{k=0}^2{(-1)^k\over (2k)!!}\,\ln^k(x^+\,x^-)\,\left(
\ell_{3-k}(x^+) - \ell_{3-k}(x^-)\right)\,,\\
\ell_n(x) &\,= {1\over 2}\left(\textrm{Li}_n(x) - (-1)^n\textrm{Li}_n(1/x)\right)\,,
\esp\eeq
and
\beq
\chi = \left\{\begin{array}{ll}
-2, & \textrm{ if } \Delta >0 \textrm{ and } u_1+u_2+u_3>1\,,\\
1, & \textrm{otherwise}\,.
\end{array}\right.
\eeq
Some comments are in order about this expression: Firstly, for $\Delta\neq0$, Eq.~\eqref{eq:result} is of manifestly uniform transcendental weight three. Secondly, we performed the computation in the Euclidean region where the integral is real and all Mandelstam invariants are taken to be negative, and hence the cross ratios are positive. Eq.~\eqref{eq:result} is proportional to the square root of $\Delta$, which becomes purely imaginary for certain values of the cross ratios in the Euclidean region. It is easy to see that in the region where $\Delta<0$, the variables $x_i^\pm$ are complex conjugate to each other, such that
\beq
\ell_n(x_i^+) - \ell_n(x_i^-) = 2i\,\textrm{Im}\left(\ell_n(x_i^+)\right)\,,
\eeq 
and hence the whole expression stays real even for $\Delta<0$. Thirdly, the branch cuts of the polylogarithms appearing in Eq.~\eqref{eq:result} are fixed by assigning a small positive imaginary part to each of the $x_i^\pm$, \emph{i.e.}, all the polylogarithms are interpreted as $\ell_n(x_i^\pm+i\varepsilon)$. In order to account for discontinuities in the $\ell_n$ functions, we need to introduce the parameter $\chi$, and the resulting expression~\eqref{eq:result} is smooth everywhere. Alternatively, we can choose $\chi = 1$ everywhere, to the price of evaluating some of the polylogarithms on a different Riemann sheet. To see this, note that in the region where $\Delta<0$, we have $\textrm{sign}(\textrm{Im}(x_i^\pm))=\pm1$ and $\chi=1$. Hence, if $\Delta$ becomes positive, $x_i^+$ ($x_i^-$)  approaches the real axis from above (below). Thus we can choose $\chi=1$ even for $\Delta>0$, but we have to evaluate the polylogarithms with a modified $i\varepsilon$ prescription, $\ell_n(x_i^\pm\pm i\varepsilon)$. Finally, let us comment on what happens at $\Delta=0$.  Eq.~\eqref{eq:result} apparently has a singularity at these points. However, it is easy to see that if $\Delta$ vanishes, the combination $\ell_n(x_i^+)- \ell_n(x_i^-)$ must vanish accordingly, and the resulting expression stays finite. In particular, it follows from Corollary 26 of Ref.~\cite{Brown:2009rc} that for $\Delta=0$ the weight of the expression must drop to less than three. As an example, $\Delta$ vanishes at $u_1=u_2=u_3=1$, and it is easy to check that in this case the weight drops by one unit,
\beq
\lim_{u_i\to1}\cI_6(u_1,u_2,u_3) = {\pi^2\over3}\,.
\eeq

\section{Conclusion}

We evaluated the massless one-loop hexagon integral in six dimensions. The result is given in terms of standard polylogarithms
of uniform transcendental weight three, its functional form resembling the one of the remainder function of the two-loop hexagon Wilson loop
in four dimensions, which has uniform weight four. Thus, it is natural to expect that one-loop hexagon integrals in six dimensions be used as a computational probe of two-loop amplitudes and Wilson loops, yet to be evaluated.

\section{acknowledgements}
The authors are grateful to Lance Dixon for useful discussions. CD is grateful to Herbert Gangl for valuable discussions on the symbol technique. 
This work was partly supported by the Research Executive Agency (REA) of the European Union through the Initial Training Network LHCPhenoNet under contract PITN-GA-2010-264564, and by the Russian Foundation for Basic Research through grant 11-02-01196.

\end{document}